\documentclass[3p,times,twocolumn]{elsarticle}

\usepackage{ecrc}

\volume{00}

\firstpage{1}

\journalname{Nuclear and Particle Physics Proceedings}

\runauth{A. Maj}

\jid{nppp}

\jnltitlelogo{Nuclear and Particle Physics Proceedings}

\usepackage{amssymb}

\def \beq  {\begin{equation}}
\def \eeq  {\end{equation}}
\def \beqar {\begin{eqnarray}}
\def \eeqar {\end{eqnarray}}
\def\CP {\mathbb{CP}}

\def\ba {\bar{a}}
\def\bb {\bar{b}}
\def\bA {\bar{A}}
\def\bD {\bar{D}}
\def\bB {\bar{B}}
\def\bC {\bar{C}}

\def\a {\alpha}
\def\b {\beta}
\def\e {\epsilon}
\def\bchi {\bar{\chi}}
\def\Tr {{\rm Tr}}
\def\dag {\dagger}
\def\bnabla {\bar{\nabla}}

\def\A {{\cal A}}

\def\C {{\cal C}}
\def\D {{\cal D}}

\def\G {{\cal G}}

\def\M{{\cal M}}
\def\O {{\cal O}}
\def\P {{\cal P}}

\def\la {{\langle}}
\def\ra {{\rangle}}

\begin{document}

\begin{frontmatter}

\dochead{}

\title{A gauge-invariant measure for gauge fields on $\CP^2$\tnoteref{a}}

\tnotetext[a]{Talk given at the 27th International Conference in Quantum Chromodynamics (QCD24), 8 - 12 July 2024, Montpellier, France.}

\author[1,2,3]{Antonina Maj}
\ead{amaj@gradcenter.cuny.edu}

\address[1]{The Graduate Center, CUNY, New York, NY 10016, USA}
\address[2]{Physics and Astronomy Department, Lehman College, CUNY, Bronx, NY 10468, USA}
\address[3]{Physics Department, City College of New York, CUNY, New York, NY 10031, USA}

\begin{abstract}
We consider four-dimensional non-Abelian gauge theory living on a complex projective space $\CP^2$ as a way of gaining insights into (3+1)-dimensional QCD.
In particular, we use a complex parametrization of gauge fields on which gauge transformations act homogeneously. 
This allows us to factor out the gauge degrees of freedom from the volume element leading to a manifestly gauge-invariant measure for the gauge-orbit space (the space of all gauge potentials modulo gauge transformations).
The terms appearing in the measure that are of particular interest are mass-like terms for the gauge-invariant modes of the gauge fields. 
Since these mass terms come with dimensional parameters they are significant in the context of dimensional transmutation.
Moreover, the existence of local gauge-invariant mass terms on $\CP^2$ could be related to Schwinger-Dyson calculations of the soft gluon mass.
Finally, we argue that there is a kinematic regime in which the theory can be approximated by a 4d Wess-Zumino-Witten (WZW) theory. This result can be used to draw similarities between the mechanism of confinement in four and (2+1) dimensions.
\end{abstract}


\end{frontmatter}


\section{Introduction}\label{intro}
The low energy or non-perturbative regime of non-Abelian gauge theories is a notoriously difficult area of research even after decades of work. Insights into phenomena ocurring at the low energy scale rely on theoretically simplified models and lattice simulations. In this paper we propose a novel contribution to the rich body of work done in non-perturbative methods for gauge theories. 
Of particular interest is the gauge-invariant volume element for the gauge fields, or the measure for the gauge-orbit space (space of gauge fields modulo gauge transformations), which is an indispensible element of the functional description of a gauge theory. In 4d non-Abelian gauge theories there is no satisfactory volume element that is manifestly gauge-invariant in the continuum limit. The measure is usually found with a gauge-fixing procedure and holds only perturbatively, such as is the case with the Faddeev-Popov or BRST procedures. 
By using a complex parametrization of gauge fields on which gauge transformations act homogenously, we are able to factor out the gauge degrees of freedom from the volume element leading to a manifestly gauge-invariant measure of integration that holds at the non-perturbative, low energy scale of the theory. Such a measure is arguably one of the most important starting points to an analytical understanding of the non-perturbative dynamics of gauge theories. This paper is an overview of the work done by Dimitra Karabali, Parameswaran Nair and myself. For a more in-depth analysis see \cite{KMN1, KMN2}.

The motivation for this project comes from (2+1)-dimensional gauge theories, where the measure for the gauge-orbit space was calculated exactly in terms of a Wess-Zumino-Witten (WZW) action using a complex parametrization of the gauge fields \cite{2d-A}. This measure was used in the Hamiltonian description of a (2+1)-dimensional gauge theory leading to a formula for the string tension (which agrees with lattice results to a remarkable degree) \cite{2d-D, 2d-E}, as well as insights into the mass gap \cite{2d-B, 2d-C}. These successful findings for a (2+1)d theory suggest that a key ingredient to understanding non-perturbative features of non-Abelian gauge theories, such as confinement and the mass gap, lies in the gauge-invariant functional measure.

The complex parametrization of gauge fields in (2+1)d is possible due to the fact that a 2d Euclidean space can be considered a complex manifold. By contrast, 4d Euclidean space has no unique complex structure. There is no unique way of assembling four real coordinates into two complex ones, and imposing a particular choice will lead to a space with fewer rotational degrees of freedom. Therefore, by considering a complex parametrization of gauge fields in four dimensions we break Euclidean invariance of the theory.\footnote{One can recover the full Euclidean symmetry group with a twistor space structure -- a topic I leave for a different paper.} This is not unlike lattice gauge theories, where the lattice breaks Euclidean global symmetries as well. The complex manifold that we consider for our gauge theory is a complex projective space $\CP^2$, which has the advantage of being a compact space with finite volume. Therefore, unlike 4d Euclidean space, it comes with a natural infrared regulator given by its finite size.

Terms of particular interest that we find in the gauge-invariant measure on $\CP^2$ are mass terms that are local in the new parametrization and manifestly gauge-invariant. These mass terms are significant in the context of discussing the possibility of gluon propagator mass. There is growing evidence from lattice simulations \cite{soft-latt1, soft-latt2, soft-latt3, soft-latt4} and analytical results from Schwinger-Dyson equations \cite{soft-an1, soft-an2, soft-an3} to the effect that a gluon "mass" emerges in the zero momentum limit of the gluon self-energy, despite the fact that an explicit mass term for gauge fields must break either Lorentz invariance, gauge invariance or locality. On $\CP^2$ we have local, gauge-invariant mass terms that could be used to shed light on the possibility of gluon propagator mass.

Moreover, given that mass terms come with dimensional parameters in a 4d theory, our results are significant in the context of dimensional transmutation. The dimensional parameters coming from the measure could provide a scale to a pure 4d Yang-Mills theory and be linked to $\Lambda_{\rm QCD}$.

Finally, among the mass terms there is a four-dimensional WZW action for Hermitian matrix-valued fields. This result brings us in direct analogy to the (2+1)d theory, where the measure was given exactly by a 2d WZW action for Hermitian fields \cite{2d-A}. In the (2+1)d theory the WZW action in the measure played a crucial role in leading to an area law behavior of the expectation value of the Wilson loop \cite{2d-D}. An analogous calculation is out of reach for the 4d case, but important insights are possible from the analogy with the (2+1)d theory.

The paper is organized as follow. In section \ref{par} I discuss the complex parametrization of the gauge fields and identify the gauge and gauge-invariant degrees of freedom. Section \ref{measure} deals with the gauge-invariant measure of integration. Finally, in section \ref{results} I state the results in the flat space limit, and section \ref{disc} is a discussion of the key results.

\section{Parametrization}\label{par}
We consider an $SU(N)$ gauge theory living on a complex projective space $\CP^2$. The metric that we use on $\CP^2$ is the standard Fubini-Study metric which is given in local coordinates $z^a$, $\bar{z}^{\ba}$, $a=1,2$, $\ba =1,2$, by
\beq
ds^2 = {dz \cdot d\bar{z} \over (1+ z\cdot \bar{z}/r^2)} - {\bar{z} \cdot dz \, z\cdot d\bar{z} \over r^2 (1+ z\cdot \bar{z} /r^2)}
\label{cp2-1}
\eeq
where $r$ parametrizes the volume of the space and serves as the infrared cutoff for the theory. The corresponding volume element is given by
\beq
d\mu = {2 \over \pi^2} {d^4 x \over (1 + z\cdot \bar{z}/r^2)^3}
\label{cp2-2}
\eeq
where the volume element is normalized so that the volume of $\CP^2$ is $r^4$. It is easy to see that in the $r \rightarrow \infty$ limit we recover flat $\mathbb{C}^2$ space ($ds^2 \rightarrow dz \cdot d\bar{z}$ and $d\mu \rightarrow {2\over \pi^2} d^4x$), which makes it easy to compare any results to lattice calculations in flat space. 

The manifold $\CP^2$ is the group coset space $SU(3)/U(2)$. This allows us to use group theory tools to identify the most general form of a vector and, hence, gauge field on $\CP^2$. In particular, coordinates on $\CP^2$ can be parametrized in terms of group elements $g \in SU(3)$, with the identification $g \sim gh$, $h \in U(2) \subset SU(3)$. The $U(2)$ subgroup defines the local isotropy group, i.e., it tells us how coordinates on $\CP^2$ behave under local rotations. Therefore, scalars, vectors and tensors will be defined by how they behave under the $U(2)$ subgroup. 

We take the Lie algebra of $SU(3)$ in the fundamental representation to be spanned by Gell-Mann matrices $\lambda_a$, with the $U(2)$ subgroup given by the directions $a = 1, 2,3$ and $a=8$, corresponding to isospin and hypercharge, respectively. Scalar functions on $\CP^2$ will then be functions of group elements $g \in SU(3)$ that are invariant under the $U(2)$ subgroup, i.e., $f(gh)=f(g)$. By looking at the gradient of a scalar function, one can extrapolate the $U(2)$ properties of a vector. This leads to the general parametrization of Abelian gauge fields on $\CP^2$ as
\beqar
A_a &=& -\nabla_a f - g_{a\ba} \e^{\ba \bb} \bnabla_{\bb} \chi \nonumber\\
\bA_{\ba} &=&\bnabla_{\ba} \bar{f} + g_{\ba a} \e^{ab} \nabla_b \bchi
\label{cp2-3}
\eeqar
where $g_{a \ba}$ is the metric tensor on $\CP^2$ from (\ref{cp2-1}), and $\e^{ab}$ and $\e^{\ba \bb}$ are antisymmetric tensors on $\CP^2$.
In the above $f$ and $\bar{f}$ are scalar fields, and $\chi$ and $\bchi$ are fields with non-trivial hypercharge ($Y = \pm 2$) (in the flat limit of $\CP^2$ $\chi$ and $\bchi$ become scalar functions).\footnote{In fact it is easy to see that $A_a$ must carry hypercharge $Y=1$, as does the derivative $\nabla_a$ (likewise, $\bA_{\ba}$ must carry $Y=-1$). Therefore, it follows that $f$ and $\bar{f}$ must have $Y=0$, and $\chi$ and $\bchi$ must have $Y=2$ and $Y=-2$, respectively.} We chose the gauge fields to be anti-Hermitian. The parametrization in (\ref{cp2-3}) can be seen as a Hodge decomposition of vector fields on $\CP^2$ (decomposition of vector fields in terms of a curl-free and a divergence-free term). The details leading to this parametrization are given in \cite{KMN1}.

In going to the non-Abelian case one can in principle simply promote the $f$ and $\chi$ fields to Lie-algebra-valued matrices. However, to get a gauge-invariant parametrization, first it is worth mentioning that multiplying a field by a scalar function does not change the field's $U(2)$ properties. Therefore, it is straightforward to show that a general non-Abelian gauge field can be parametrized as
\beqar
A_a &=& -\nabla_a M M^{-1} -M^{\dag -1} (g_{a \ba} \e^{\ba \bb} \bnabla_{\bb} \chi ) M^{\dag} \nonumber\\
\bA_{\ba} &=& M^{\dag -1} \bnabla_{\ba} M^{\dag} + M (g_{\ba a} \e^{ab} \nabla_b \bchi ) M^{-1}
\label{cp2-4}
\eeqar
where $M$ and $M^{\dag}$ are elements of the complexification of the gauge group, $M, M^{\dag} \in SL(N, \mathbb{C})$, and they are matrix-valued scalar fields. $\chi$ and $\bchi$ are Lie-algebra-valued fields that have the same non-trivial hypercharge as in the Abelian case.

Under a gauge transformation $U \in SU(N)$ it is easy to see that $M \rightarrow U M$, $M^{\dag} \rightarrow M^{\dag} U^{\dagger}$, and $\chi$ and $\bchi$ remain invariant. For any square complex matrix we can perform a polar decomposition of the form $M = U \rho$, where $U$ is unitary and $\rho$ is Hermitian. A gauge transformation will only affect the unitary part of $M$ and $M^{\dag}$. Therefore, we conclude that the gauge-invariant degrees of freedom are given by a Hermitian scalar field $\rho$, and the $\chi$ and $\bchi$ fields. Instead of $\rho$ we will be using $H \equiv M^{\dag} M = \rho^2$ as the Hermitian degree of freedom.

It is useful to re-express the gauge fields in (\ref{cp2-4}) as
\beqar
A_a &=& -\nabla_a M M^{-1} - M a_a M^{-1} \nonumber\\
\bA_{\ba} &=& M^{\dag -1} \bnabla_{\ba} M^{\dag} + M^{\dag -1} \ba_{\ba} M^{\dag}
\label{cp2-5}
\eeqar
where $a_a = g_{a \ba} \e^{\ba \bb} H^{-1} \bnabla_{\bb} \chi H$ and $\ba_{\ba} = g_{\ba a} \e^{ab} H \nabla_b \bchi H^{-1}$. It is easy to see that $a_a$ and $\ba_{\ba}$ obey the conditions $\bar{\D} \cdot a = \D \cdot \ba = 0$, where $\D$ and $\bar{\D}$ are covariant derivatives with the connections $-\nabla H H^{-1}$ and $H^{-1} \bnabla H$, respectively. Therefore, $a_a$ and $\ba_{\ba}$ are not four independent fields but two (times the dimension of the gauge group), as are $\chi$ and $\bchi$. Similarly as $\chi$ and $\bchi$, $a_a$ and $\ba_{\ba}$ are gauge-invariant. 

Finally, it is worth mentioning one more feature. The parametrization in (\ref{cp2-4}) (or (\ref{cp2-5})) is complete in the sense that any gauge field can be written in this way. However, it is not unique. A choice of parameters $M \bar{V} (\bar{x})$, $V(x) M^{\dag}$, $V(x) \chi V(x)^{-1}$, $\bar{V}(\bar{x})^{-1} \bchi \bar{V}(\bar{x})$, where $V$ and $\bar{V}$ are holomorphic and anti-holomorphic scalar matrices, leads to the same gauge fields. This holomorphic ambiguity of the parametrization could be useful in the context of the Gribov problem on $\CP^2$. Since $\CP^2$ has a Gribov problem \cite{Grib}, the parametrization given by $H$, $\chi$ and $\bchi$ cannot globally parametrize the gauge-orbit space, but can only hold locally on coordinate patches. If so, then we must find appropriate transition functions that take us from one patch to another. In fact this kind of analysis was done for (2+1)d gauge theories \cite{2d-A}, where gauge fields were parametrized in terms of $M$ and $M^{\dag}$ alone. There the transition functions on the gauge-orbit space were given precisely by the holomorphic and anti-holomorphic functions $V$ and $\bar{V}$. Therefore, in the (2+1)d case a measure that is invariant under these (anti)holomorophic transformations is invariant under the transition functions and, so, the measure holds globally on the gauge-orbit space. An analogous analysis for $\CP^2$ is left for future investigation.

\section{Measure}\label{measure}
To find the gauge-invariant measure for the gauge fields we first write the measure in terms of our new parameters $M$, $M^{\dag}$, $\chi$ and $\bchi$, and then factor out the volume element corresponding to the unitary, or gauge, part of $M$ and $M^{\dag}$. We start by considering the metric on the space of gauge fields on $\CP^2$,
\beq
ds^2 = - \int d\mu \, g^{a\ba} \, \Tr (\delta A_a \, \delta \bA_{\ba}) 
\label{cp2-6}
\eeq
where $d\mu$ is the volume element on $\CP^2$ given in (\ref{cp2-2}). Using the parametrization in (\ref{cp2-4}) and (\ref{cp2-5}),
\beqar 
\delta A_a &= - D_a \theta - g_{a\ba}\e^{\ba\bb} \bD_{\bb} \delta \tilde{\chi} + [\theta^{\dag}, Ma_a M^{-1}] \nonumber\\
\delta \bA_{\ba} &= \bD_{\ba} \theta^{\dag} + g_{\ba a} \e^{ab} D_b \delta \tilde{\chi}^{\dag} + [\theta, M^{\dag-1} \ba_{\ba} M^{\dag}]
\label{cp2-7}
\eeqar
where $\theta =\delta M M^{-1}$, $\theta^{\dag} = M^{\dag -1} \delta M^{\dag}$, $\delta \tilde{\chi} = M^{\dag -1} \delta \chi M^{\dag}$, and $\delta \tilde{\chi}^{\dag} = M \delta \bchi M^{-1}$. The covariant derivatives $D$ and $\bD$ have $-\nabla M M^{-1}$ and $ M^{\dag -1} \bnabla M^{\dag}$ as their connections, respectively. 
Using (\ref{cp2-7}) in (\ref{cp2-6}) we can write the metric as a quadratic form
\beq
ds^2 = {1 \over 2} \int d\mu \, \xi^{\dag \a} \M_{\a \b} \xi^{\b}, \hskip.1 in  \xi = (\theta, \theta^{\dag}, \delta \tilde{\chi}, \delta \tilde{\chi}^{\dag})
\label{cp2-8}
\eeq
where $\xi^{\a}$ is written in terms of Lie algebra components by taking $\xi = - i t_{\a} \xi^{\a}$, and $\M_{\a\b}$ is in the adjoint representation of the gauge group. $\M$ is effectively the Jacobian matrix of the coordinate transformation $(A_a, \bA_{\ba}) \rightarrow (M, M^{\dag}, \chi, \bchi)$.

The volume element corresponding to the metric in (\ref{cp2-8}) is given by
\beq
dV = \sqrt{\det \M} \, dV_0
\label{cp2-9}
\eeq
where $dV_0$ is the volume element corresponding to the metric for $\xi$
\beqar
ds_0^2 &=& {1\over 2} \int d\mu \xi^{\dag \a} \xi^{\a} = \int d\mu ( \theta^{\dag \a}\theta^{\a} + \delta \tilde{\chi}^{\dag\a} \delta\tilde{\chi}^{\a}) \nonumber\\
&=& \int d\mu \left[ (M^{\dag-1}\delta M^{\dag})^{\a} (\delta M M^{-1})^{\a}  \right. \nonumber\\
&& \hskip .35in \left. + \, \delta \bchi^{\a} H_{\a \b} \delta \chi^{\b} \right]
\label{cp2-10}
\eeqar
In the second term $H_{\a\b}$ is the adjoint representation of $H$ defined by $H_{\a\b} = 2\Tr (H t_\a H^{-1} t_\b)$. Given that $H$ has unit determinant, the second term in (\ref{cp2-10}) corresponds to the measure for the $\chi$ and $\bchi$ fields, $ [d\chi d\bchi]$. The first term in (\ref{cp2-10}) is the Cartan-Killing metric for $SL(N, \mathbb{C})$ and the corresponding volume element is the Haar measure, $d\mu (M, M^{\dag})$. By using a polar decomposition of $M$ and $M^{\dag}$ it is easy to show that the Haar measure for $SL(N, \mathbb{C})$ decomposes into a Haar measure for the $SU(N)$ subgroup, $d\mu (U)$, and a Haar measure for $SL(N, \mathbb{C})/SU(N)$, $d\mu(H)$. Therefore, the volume element for gauge fields in the new parametrization is given by
\beq
dV = \sqrt{\det \M} \, [d\mu(U) d\mu(H) d\chi d\bchi]
\label{cp2-11}
\eeq
To get the gauge-invariant measure we simply factor out the volume for the gauge transformations, i.e., $d\mu(U)$, leading to
\beq
dV [\C] = \sqrt{\det \M} \, [d\mu(H) d\chi d\bchi]
\label{cp2-12}
\eeq
where $\C = \A / \G_*$ denotes the gauge-orbit space, the space of all gauge fields ($\A$) modulo gauge transformations ($\G_*$).

The final step towards getting the gauge-invariant measure is the calculation of the determinant of the matrix $\M$ in (\ref{cp2-8}). The easiest way to do this is by writing the determinant as a functional integral over auxiliary bosonic fields $B, \bB, C, \bC$,
\beq
{1 \over \sqrt{\det \M}} = \int [dB d\bB dC d\bC] e^{-S_0 - \Delta S}
\label{cp2-13}
\eeq
where
\beqar
S_0 &=& \int d\mu [ \bC^{\a} (-\bD \cdot D)_{\a \b} C^\b \nonumber\\
&& \hskip.4in+B^\a (-\bD \cdot D)_{\a\b} \bB^{\b}] \label{cp2-14}\\
\Delta S &=& \int d\mu \left[\bC^{\a} (MaM^{-1}\cdot M^{\dag -1} \ba M^{\dag})_{\a\b} C^\b \right. \nonumber\\
&& \hskip.4in +C^{\a} (M^{\dag -1} \ba M^{\dag} \cdot D )_{\a\b} C^{\b} \nonumber\\
&&\hskip .4in + \bC^{\a} (-M a M^{-1} \cdot \bD)_{\a\b} \bC^{\b} \nonumber\\
&& \hskip .4in +C^\a (-\e^{\ba\bb} M^{\dag -1} \ba_{\ba}M^{\dag} \bD_{\bb})_{\a\b} B^{\b} \nonumber\\
&& \hskip .4in \left.+ \bC^{\a} (\e^{ab} M a_a M^{-1} D_b) \bB^{\b} \right]
\label{cp2-15}
\eeqar
In the above $C$ and $\bC$ behave like $\theta$ and $\theta^{\dag}$, i.e., like scalars, and $B$ and $\bB$ carry the same hypercharge as $\chi$ and $\bchi$, respectively. We calculate the integral in (\ref{cp2-13}) as a perturbation series in powers of $\Delta S$, and hence in powers of $a$ and $\ba$. Therefore,
\beq
\sqrt{\det \M} = e^{\Gamma}, \hskip .2in \Gamma = \Gamma_0 +\Delta \Gamma
\label{cp2-16}
\eeq
where
\beqar
\Gamma_0 &=& \Tr \log(-\bD \cdot D)_{Y=0} + \Tr \log (-\bD \cdot D)_{Y=-2} \nonumber\\
\Delta \Gamma &=& - \log \left( { \int e^{- S_0 - \Delta S} \over \int e^{-S_0}}\right)
\label{cp2-17}
\eeqar
where the subscript $Y=0$ denotes the hypercharge of a scalar field, and $Y=-2$ denotes the hypercharge of a $\bchi$ field. 

The remaining element left for calculating $\Gamma$ are the propagators for scalar and $Y=-2$ fields, $\G = (-\bD \cdot D)^{-1}$. For this we first calculate the free propagators (propagators in the absence of gauge fields, $(-\bnabla \cdot \nabla)^{-1}$) for both scalar and $Y=-2$ fields. We then expand $\G$ in powers of $\nabla M M^{-1}$ and $M^{\dag -1} \bnabla M^{\dag}$. Finally, we include a covariant point-splitting type regulator for the UV divergencies that is both gauge-covariant and respects the underlying symmetries of $\CP^2$. As mentioned above, since $\CP^2$ is a space of finite volume, we need not worry about IR divergencies. These calculations come with multiple subtleties that go beyond the scope of this paper. I refer the interested reader to \cite{KMN1, KMN2} for more details.

Finally, it is worth commenting on the structure of $\Gamma$ in (\ref{cp2-16}). In general, $\Gamma$ will have terms of increasing scaling dimension. Terms of lowest scaling dimension will be terms of dimension 2 that can in principle be quadratically UV divergent. There will also be terms of scaling dimension 4 that will be logarithmically UV divergent. And, finally, there will be further terms of higher dimensions that will be UV finite that will carry more and more powers of $r^2$, where $r$ parametrizes the volume of $\CP^2$. In the flat limit, $r \rightarrow \infty$, these terms become increasingly infrared divergent. We expect the terms of lowest scaling dimension to be dominant in the low energy regime of the theory and, therefore, we want a way to parametrically control the further finite terms. We do this by introducing an infrared cutoff $\lambda$. We then calculate loop diagrams for momenta ranging from $\lambda$ up to a UV cutoff, effectively integrating out modes with momenta higher than $\lambda$. The resulting theory will be an effective theory for low energy fields. UV finite terms that carried increasing powers of $r^2$ will now carry increasing inverse powers of $\lambda$. For fields with momenta much smaller than $\lambda$ these further terms will be of order $\O (k^2 / \lambda)$ and, therefore, they will be subdominant compared to the terms of lowest scaling dimension. Hence, we are only interested in the first few terms that can potentially be UV divergent -- terms of scaling dimension 2 and 4. 

To see how this works in practice, in the next section I will consider the flat limit of $\CP^2$ as a way of simplifying calculations. This is also the most relevant case for QCD. Readers interested in the theory on the full $\CP^2$ space should refer to \cite{KMN1, KMN2}.

\subsection{Regularizations}\label{reg}
As mentioned above, in the flat limit $\CP^2$ becomes $\mathbb{C}^2$. Moreover, the $\chi$ and $\bchi$ fields that on $\CP^2$ carry a non-trivial hypercharge in the flat limit become scalar fields. The free propagator $G=1/(-\bnabla \cdot \nabla)$ for $\mathbb{C}^2$ is given by\footnote{For 4d space the Greens function for $-\nabla^2$ is usually given by $G(x,y) = {1\over \pi^2 |x-y|^2}$. The normalization we use here comes from the fact that on $\CP^2$ we use a measure that normalizes the volume of $\CP^2$ to $r^4$, as given in (\ref{cp2-2}). In the flat limit this measure is equivalent to $d\mu(\CP^2) \rightarrow {2 \over \pi^2} d^4 x$.}
\beq
G(x,y)=\left({1 \over - \bnabla \cdot \nabla}\right)_{x,y}= {1 \over 2|x-y|^2}
\label{cp2-18}
\eeq
We regularize the UV divergencies with a point-splitting type regulator. In the propagator $G(x,y)$ we move $y$ to a nearby point $y' = y + \delta y$, where $\delta y^a \delta \bar{y}^{\ba} \rightarrow \e \delta^{a \ba}$ in taking the small $\e$ limit in a symmetric way. Effectively, this means that 
\beq
G_{\rm Reg}(x,y)=G(x, y') \rightarrow {1\over 2 (|x-y|^2 + \e)}
\label{cp2-19}
\eeq

In the presence of gauge fields the propagator of interest is $\G = 1/(-\bD \cdot D)$, where $D$ has $-\nabla M M^{-1}$ as its connection and $\bD$ has $M^{\dag-1}\bnabla M^{\dag}$. Under a gauge transformation $M \rightarrow UM$ and $M^{\dag} \rightarrow M^{\dag} U^{\dag}$. Therefore, the full propagator must transform as $\G(x,y) \rightarrow U(x) \G(x,y) U^{\dag}(y)$. When regulating the propagator by point-splitting as described above, we must make sure that it transforms in the same way as the unregulated propagator. For this purpose we include a Wilson line for the connections that connects $y'$ to $y$. Hence, the gauge-covariant point splitting regularization is
\beqar
\G_{\rm Reg}(x,y) &=& \G(x,y') W(y',y) \label{cp2-20}\\
W(y',y) &=& \P \exp \left( \int_y^{y'}\nabla M M^{-1} - M^{\dag -1} \bnabla M^{\dag} \right) \nonumber
\eeqar
where $\G$ is expanded in $-\nabla M M^{-1}$ and $M^{\dag -1} \bnabla M^{\dag}$ fields in the following way
\beq
\G(x,y) = G(x,y) + \int_z G(x,z) \mathbb{X}(z) G(z,y) +\cdots
\label{cp2-20b}
\eeq
$\mathbb{X}$ is defined by $-\bD\cdot D \equiv -\bnabla \cdot \nabla - \mathbb{X}$, and given by $\mathbb{X} = M^{\dag -1} \bnabla M^{\dag} \cdot \nabla - \nabla M M^{-1} \cdot \bnabla - \bnabla ( \nabla M M^{-1}) - M^{\dag -1} \bnabla M^{\dag} \cdot \nabla M M^{-1}$.

As discussed in the previous section we also include an IR cutoff as a way of parametrically controlling the finite terms in the effective action $\Gamma$. In the flat limit the IR cutoff is needed for another reason, because the finiteness of the volume of $\CP^2$ can no longer serve as an IR regulator. We include the IR cutoff $\lambda$ through a simple integral representation of the free propagators, where we include $\lambda$ as the lower cutoff
\beq
G(x,y) = \int_{\lambda}^{\infty} {ds\over2} \, e^{-s|x-y|^2}
\label{cp2-21}
\eeq
When $\lambda$ is set to zero we clearly reproduce the free propagator in (\ref{cp2-18}).

\section{Results}\label{results}
Having discussed the propagator and its regulators one can in principle calculate the gauge-invariant measure in (\ref{cp2-12}). The results are as follows
\beq
dV[\C] = e^{\Gamma(H,\chi,\bchi)} [d\mu(H) \, d\chi d\bchi ]
\label{cp2-22}
\eeq
where in the flat space limit $\Gamma$ is given by 
\beqar
\Gamma &=& {\lambda \over 2 \pi} S_{\rm wzw} (H) -{1 \over 4 \e} \int \Tr(\ba \cdot H a H^{-1}) \nonumber\\
&& + {\log \e \over 12} \int \Tr \left[(\bnabla(\nabla H H^{-1}))^2 + (\ba \cdot H a H^{-1})^2 \right. \nonumber\\
&& \hskip .6in + [\ba,HaH^{-1}] \bnabla(\nabla H H^{-1}) \nonumber \\
&& \hskip.6in - g^{a\ba} g^{b \bb} \bnabla_{\ba}(\nabla_bH H^{-1}) [ \ba_{\bb}, H a_a H^{-1}] \nonumber\\
&&\hskip .6in \left.- g^{a\ba} g^{b\bb} \bnabla_{\ba} \ba_{\bb} \D_a(Ha_bH^{-1}) \right] \nonumber\\
&& + \O(k^2 / \lambda)
\label{cp2-23}
\eeqar
The first term $S_{\rm wzw}(H)$ is the four-dimensional WZW action given by
\beqar
S_{\rm wzw}(H) &=& {\pi \over 4} \int d\mu \,\Tr (\nabla H \bnabla H^{-1})  \nonumber\\
&&  - {i \over 24 \pi} \int \omega \wedge \Tr (H^{-1} d H )^3
\label{cp2-24}
\eeqar
where $\omega$ is the K\"ahler two-form on $\CP^2$ given in local coordinates by $\omega = i g_{a\ba} dz^a d\bar{z}^{\ba}$. The second term in (\ref{cp2-24}) is an integral over a five-manifold which has $\CP^2$ as the boundary.

The first two terms in (\ref{cp2-23}) are of scaling dimension 2 and as such they are gauge-invariant mass terms for the $H$, and $a_a$ and $\ba_{\ba}$ modes. On grounds of dimensionality they could both be UV divergent, however, the WZW term comes with a finite coefficient that depends on the IR cutoff $\lambda$.\footnote{The WZW term is also present in the full $\CP^2$ calculation and also comes with a finite, rather than UV divergent, coefficient.}

The log-divergent terms are of scaling dimension 4. Unlike in the case of Euclidean space, these terms do not combine into $\Tr F^2$, where $F$ is the field strength tensor. The reduced isometries of $\CP^2$ (or $\mathbb{C}^2$), as compared to Euclidean space, allows for additional tensor structures on top of $\Tr F^2$. Presumably, to make the volume element well-defined, counterterms of the same form as in (\ref{cp2-23}) have to be introduced and renormalization has to be carried out.

Further terms indicated by $\O (k^2/ \lambda)$ are UV finite terms of scaling dimension $>4$. As such they will go as increasing powers of $k^2/\lambda$, where $k^2$ is the momentum of the fields. We ignore these terms as they are subdominant in the low energy limit given by $k^2 \ll \lambda$.

\section{Discussion}\label{disc}
To conclude, we found a complex parametrization of a 4d non-Abelian gauge theory on $\CP^2$ on which the gauge transformations act homogenously (see section \ref{par} and equation (\ref{cp2-5}), in particular). We then used this parametrization to obtain the manifestly gauge-invariant measure of integration for the gauge fields. The measure is given by
\beq
dV[\C] = e^{\Gamma} \, [d\mu(H) \, d\chi d\bchi ]
\label{cp2-25}
\eeq
where, unlike for the (2+1)-dimensional theory (where $\Gamma$ is given exactly in terms of a WZW action \cite{2d-A}), for the four-dimensional theory we cannot find $\Gamma$ exactly. Instead, we have calculated the first few terms of scaling dimension $\leq 4$ (the results for flat space are given in equation (\ref{cp2-23}), for results for the full $\CP^2$ theory see \cite{KMN1, KMN2}). These terms are potentially UV divergent and, therefore, they are important in understanding the nature of the counterterms needed for carrying out renormalization. Furthermore, these terms are the most dominant in the low energy limit of the theory defined through the introduction of an infrared cutoff $\lambda$ (for details see subsection \ref{reg}). Therefore, in discussing the physical relevance of the results I will focus on the first two terms in (\ref{cp2-23}) that are of lowest scaling dimension and, hence, are the most dominant terms in the low energy limit of the theory. After introducing appropriate counterterms and renormalizing, the terms of interest are
\beq
\Gamma = {\lambda \over 2 \pi} S_{\rm wzw} (H) -{\mu_{\rm ren}^2} \int \Tr (\ba \cdot H a H^{-1}) +\cdots 
\label{cp2-26}
\eeq
where the elipsis includes terms of higher scaling dimension (renormalized log-divergent and UV finite terms). 

The first term is a 4d WZW action for the Hermitian $H$ fields. On dimensional grounds, this term could be quadratically UV divergent, but surprisingly it comes with a finite coefficient. The second term is UV divergent and, therefore, the measure has to be defined by introducing a counterterm for it resulting in an overall finite coefficient $\mu_{\rm ren}^2$. Both $\sqrt{\lambda}$ and $\mu_{\rm ren}$ have the dimensions of mass and, therefore, they can be seen as setting the mass scales for the theory. 

One area in which this is useful is dimensional transmutation. As is well-known, a four-dimensional gauge theory is scale-invariant -- the coupling constant in such a theory is dimensionless. Therefore, a scale factor needs to be introduced to make the theory well-defined. For QCD this is usually done through the one-loop correction to the running coupling constant, where $\Lambda_{\rm QCD}$ is introduced as the IR cutoff. However, our theory has dimensional parameters ($\lambda$ and $\mu_{\rm Ren}^2$) from the gauge-invariant measure for the gauge fields. Hence, there is no need to introduce further scale factors. In fact, if one-loop calculations were carried out in our theory, we expect $\lambda$ and $\mu_{\rm Ren}^2$ to enter as the IR cutoffs. We could, therefore, find $\Lambda_{\rm QCD}$ as a function of the parameters $\lambda$ and $\mu_{\rm Ren}^2$.

Another area for which our results are relevant is the possibility of soft gluon mass. Soft gluon mass is a dynamical, momentum-dependent mass that a gluon acquires due to self-interactions in the low momentum limit. There is considerable evidence from lattice simulations to the effect that the gluon propagator in the Landau gauge saturates at a finite non-zero value in the low momentum limit \cite{soft-latt1, soft-latt2, soft-latt3, soft-latt4}. On the analytical side, there have been calculations via Schwinger-Dyson equations that indicate that the gluon self-energy does acquire a mass in the zero momentum limit \cite{soft-an1, soft-an2, soft-an3}.

The subtlety behind the possibility of gluon mass is the fact that in Euclidean space a mass term for gauge fields either breaks the underlying symmetries of the theory or is non-local (and, importantly, cannot be made local through a suitable parametrization of the gauge fields). For example, a mass term of the form $\int \Tr A^2$ breaks gauge-invariance. To restore gauge symmetry one needs to include further non-local terms. A gauge-invariant mass term would, therefore, be of the form 
\beq
\int \Tr A^2 +\cdots \sim \int \Tr \, f^{\mu \nu} \left({1 \over -\nabla^2}\right) f_{\mu\nu} \,+\, \O(A^3)
\label{cp2-27}
\eeq
where $f_{\mu\nu} \equiv \nabla_\mu A_\nu - \nabla_\nu A_\mu$. The elipsis on the left hand side is an infinite series of non-local terms needed to ensure gauge-invariance. 

On the other hand, on $\CP^2$ the terms in (\ref{cp2-26}) are mass terms for the gauge fields that are manifestly gauge-invariant and local in the parametrization given by $H$, $a$ and $\ba$. If one were to rewrite them in terms of our original gauge fields $A_a$ and $\bA_{\ba}$, they would be non-local similarly as is (\ref{cp2-27}) for Euclidean space. However, where for Euclidean space there is no parametrization of the gauge fields that makes such a mass term local, on $\CP^2$ (and $\mathbb{C}^2$) we do have such a parametrization. Mass terms as given in (\ref{cp2-26}) could be included in a 4d Yang-Mills theory through a mass gap equation calculation analogously to what was done in the Nambu-Jona-Lasinio model \cite{NJL} and magnetic mass calculations for (2+1)-Yang-Mills theory in \cite{magn-mass}. Subsequently, we could calculate the gluon self-energy and verify whether it saturates at a non-zero value in the zero momentum limit (here, we expect this value to be a function of $\lambda$ and $\mu_{\rm Ren}^2$). Such a calculation is left for future investigation.

Other than in the context of soft gluon mass, our theory in the flat limit could be used to corroborate recent lattice results done by Chernodub et al. for the Casimir energy in a (3+1)d Yang-Mills theory \cite{chern}. They fit the Casimir energy for the gauge theory with a massive scalar field theory. On $\CP^2$ we can expand the Hermitian field $H$ near identity, $H = e^{\phi} \approx 1 + \phi$. Then, including only terms quadratic in the $\phi$, $\chi$ and $\bchi$ fields in $\Gamma$ we get a free theory for massive scalar fields. The mass terms for the scalars come from the terms in (\ref{cp2-26}). The calculations still need to be done in more detail, however, this suggests that our theory could provide an explanation for the scalar theory fit of the Casimir energy for (3+1)d gauge theory.

Finally, given that $\mu_{\rm Ren}$ sets the scale for the massive gauge modes $a$ and $\ba$, we can imagine a kinematic regime of momenta $k^2 \ll \mu_{\rm Ren}^2$ in which the $a$ and $\ba$ modes are not created. In this limit our gauge potentials are approximated by
\beq
A \simeq - \nabla M M^{-1} , \hskip .3in \bA \simeq M^{\dag -1} \bnabla M^{\dag}
\label{cp2-28}
\eeq
and the measure is given by 
\beq
\Gamma \simeq {\lambda \over 2 \pi} S_{\rm wzw} (H) + C \int \Tr F^2(H)
\label{cp2-29}
\eeq
where the second term comes from the log-divergent H-dependent term in (\ref{cp2-23}). In particular, $\Tr F^2 (H) \sim \Tr (\bnabla (\nabla H H^{-1}))^2$, and $C$ is the renormalized value after the $\log \e $ divergence in (\ref{cp2-23}) is eliminated through a counterterm. Moreover, the dominant term in the low energy limit is the term with lowest scaling dimension, i.e., the WZW action. Therefore, we conclude that there is an energy regime of our theory in which it can be approximated by a 4d WZW theory for the fields $H \in  G^{\mathbb{C}} / G = SL(N, \mathbb{C}) /SU(N)$, where $G$ is the gauge group.

The 4d WZW theory has a rich mathematical and physical structure. It was first considered by Simon Donaldson in the 1980s in the context of holomorphic vector bundles \cite{don}. It also appears in K\"ahler-Chern-Simons theory \cite{KCS1, KCS2}, which is a generalization of Chern-Simons theory in (2+1) dimensions to 4 dimensions. 4d WZW theories also occur in the context of higher dimensional quantum Hall systems \cite{qHall}, as well as in the context of holomorphic field theories on twistor space \cite{costello}. Moreover, the equations of motion of a 4d WZW theory are antiself-dual instanton equations. Therefore, the low energy dynamics of a 4d WZW theory are dominated by instantons. This is just to show that the 4d WZW theory in itself is a rich and interesting area of research.

Another reason to think that the approximation in (\ref{cp2-29}) is significant comes from the discussion of confinement in the (2+1)d theory as analyzed in \cite{2d-D}. In (2+1)d gauge theory the measure is given exactly by a 2d WZW action for Hermitian fields $H \in SL(N,\mathbb{C})/SU(N)$. This result was used to show that the expectation value of the Wilson loop has an area law behavior. Furthermore, they were able to get a result for the string tension that agrees with lattice simulations to a remarkable degree \cite{2d-E}. In particular, they found that 
\beqar
\la W(C) \ra &=& \la \Tr \, \P e^{\oint_C \nabla H H^{-1} } \ra \nonumber\\
&=& \int \, e^{2c_A \, S^{(2d)}_{\rm wzw} (H) - {1 \over 2 e^2} \int \Tr F^2(H)}  \, \Tr\, \P e^{ \oint_C \nabla H H^{-1} } \nonumber\\
&\sim& e^{ -\sigma_R \, {\rm Area}(C)}, \hskip .3in \sigma_R = e^4 {c_A c_R \over 4 \pi}
\label{cp2-30}
\eeqar
where $S^{(2d)}_{\rm wzw} (H)$ is the 2d WZW action for $H$. $c_R$ and $c_A$ are the quadratic Casimir operator values for the representation $R$ and for the adjoint representation, respectively. $e^2$ is the coupling constant of the Yang-Mills theory.

In the approximation of our theory in (\ref{cp2-29}) there is a WZW action and a Yang-Mills action for the Hermitian fields $H \in SL(N,\mathbb{C}) / SU(N)$, similarly as in the (2+1)-dimensional case. Thus, one can in principle proceed by analogy to the (2+1)d theory to see whether the expectation value of the Wilson loop has an area law behavior. However, for the four-dimensional case an analysis as in (\ref{cp2-30}) is calculationally too complicated at this stage. We can, however, make a less rigorous argument. 

In the limit of $e^2 \rightarrow \infty$ in (\ref{cp2-30}), the Wilson loop expectation value vanishes for any curve $C$ with non-zero area. This limit also defines the 2d WZW theory approximation, as in this limit ${1 \over 2e^2}\Tr F^2 \rightarrow 0$, leaving just the WZW action. One can trace the origin of the vanishing of the Wilson loop to a UV singularity of the 2-point correlator $\oint \oint \la \nabla H H^{-1} \nabla H H^{-1} \ra$ in the absence of the Yang-Mills term. 

In the 4d case there is a similar UV singularity in the 2-point function in the absence of the $\Tr F^2$ term. Therefore, similarly as for the (2+1)d theory, in the 4d WZW theory approximation the Wilson loop vanishes for any curve with non-zero area. The difficulty, however, lies in showing that the presence of the Yang-Mills action regulates this singularity to produce an area-law behavior. A more detailed version of this argument is given in \cite{KMN2}.


\section*{Acknowledgements}
This research was supported in part by the U.S. National Science Foundation Grants No. PHY-2112729 and No. PHY-1915053, PSC-CUNY grants and by the Bernard B. Levine Graduate Fellowship.

\bibliographystyle{elsarticle-num}

\begin{thebibliography}{23}
\bibitem{KMN1}
D. Karabali, A. Maj, V.P. Nair, Gauge and Scalar Fields on $\mathbb{CP}^2$: A Gauge-invariant Analysis I, Phys. Rev. D106 (2022) 085012.

\bibitem{KMN2}
D. Karabali, A. Maj, V.P. Nair, Gauge and Scalar Fields on $\mathbb{CP}^2$: A Gauge-invariant Analysis II, Phys. Rev. D106 (2022) 085013.

\bibitem{2d-A}
D. Karabali, V.P. Nair, A gauge-invariant Hamiltonian analysis for non-Abelian gauge theories in (2+1) dimensions, Nucl. Phys. B464 (1996) 135-152.

\bibitem{2d-D}
D. Karabali, C. Kim, V.P. Nair, On the vacuum wavefunction and string tension of Yang-Mills theories in (2+1) dimensions, Phys. Lett. B434 (1998) 103-109.

\bibitem{2d-E}
D. Karabali, V.P. Nair, The robustness of the vacuum wave function and other matters for Yang-Mills theory, Phys. Rev. D77 (2008) 025014.

\bibitem{2d-B}
D. Karabali, V.P. Nair, On the origin of the mass gap for non-Abelian gauge theories in (2+1) dimensions, Phys. Lett. B379 (1996) 141-147.

\bibitem{2d-C}
D. Karabali, V.P. Nair, Gauge invariance and mass gap in (2+1)-dimensional Yang-Mills theory, Int. J. Mod. Phys. A12 (1997) 1161-1172.

\bibitem{soft-latt1}
I.L. Bogolubsky, E. M. Ilgenfritz, M. Muller-Preussker, A. Sternbeck, Lattice gluodynamics computation of Landau-gauge Green's functions in the deep infrared, Phys. Lett. B676 (2009) 69-73.

\bibitem{soft-latt2}
P.O. Bowman, et al., Scaling behavior and positivity violation of the gluon propagator in full QCD, Phys. Rev. D76 (2007) 094505.

\bibitem{soft-latt3}
O. Oliveira, P.J. Silva, The lattice infrared Landau gauge gluon propagator: The infinite volume limit, PoS LAT 2009 (2009) 226.

\bibitem{soft-latt4}
A. Cucchieri, T. Mendes, What's up with IR gluon and ghost propagators in Landau gauge? A puzzling answer from huge lattices, PoS LAT 2007 (2007) 297.

\bibitem{soft-an1}
J.M. Cornwall, J. Papavassiliou, D. Binosi, The Pinch Technique and its Application to Nonabelian Gauge Theories, Cambridge University Press 2011.

\bibitem{soft-an2}
N. Vandersickel, D. Zwanziger, The Gribov problem and QCD dynamics, J. Phys. Rep. 520(4) (2012) 175-251.

\bibitem{soft-an3}
A.C. Aguilar, D. Binosi, J. Papavassiliou, The gluon mass generation mechanism: A concise primer, Front. Phys. 11(2) (2016) 111203.

\bibitem{Grib}
T.P. Killingback, E.G. Rees, Topology of Gauge Theories on Compact Four Manifolds, Class. Quant. Grav. 4 (1987) 357.

\bibitem{NJL}
Y. Nambu, G. Jona-Lasinio, Dynamical Model of Elementary Particles Based on an Analogy with Superconductivity, Phys. Rev. 122 (1961) 345.

\bibitem{magn-mass}
G. Alexanian, V.P. Nair, A Self-consistent Inclusion of Magnetic Screening for the Quark-Gluon Plasma, Phys. Lett. B352 (1995) 435-439.

\bibitem{chern}
M.N. Chernodub, V.A. Goy, A.V. Molochkov, A.S. Tanashkin, Boundary states and Non-Abelian Casimir effect in lattice Yang-Mills theory, Phys. Rev. D108 (2023) 014515.

\bibitem{don}
S.K. Donaldson, Anti self-dual Yang-Mills connections over complex algebraic surfaces and stable vector bundles, Proc. London Math. Soc. (3) 50 (1985) 1.

\bibitem{KCS1}
V.P. Nair, J. Schiff, A K\"ahler-Chern-Simons Theory and Quantization of Instanton Moduli Spaces, Phys. Lett. 246B (1990) 423.

\bibitem{KCS2}
V.P. Nair, J. Schiff, K\"ahler-Chern-Simons Theory and Symmetries Of Anti-Self-Dual Gauge Fields, Nucl. Phys. B371 (1992) 329.

\bibitem{qHall}
D. Karabali, V.P. Nair, Edge states for quantum Hall droplets in higher dimensions and a generalized WZW model, Nucl. Phys. B697 (2004) 513.

\bibitem{costello}
K.J. Costello, Quantizing local holomorphic field theories on twistor space, arXiv:2111.08879.


\end{thebibliography}

\end{document}